\documentclass[12pt,onecolumn,journal]{IEEEtran}

\usepackage{cuted}
\usepackage{mathrsfs,amssymb,amsmath,latexsym,array, arydshln, version, bm} 
\usepackage{cite}
\usepackage{graphicx, epsfig, epic, eepic, color, subfigure}
\usepackage{setspace}
\newtheorem{theorem}{Theorem}
\newtheorem{problem}{Problem}

\newtheorem{definition}{Definition}
\newtheorem{lemma}{Lemma}
\newtheorem{proposition}{Proposition}

\def\BibTeX{{\rm B\kern-.05em{\sc i\kern-.025em b}\kern-.08em
    T\kern-.1667em\lower.7ex\hbox{E}\kern-.125emX}}
\renewcommand\IEEEQEDclosed{$\Box$}

\allowdisplaybreaks     

\begin{document}
\bibliographystyle{jcn}
\vspace{5mm}
\title{Joint Transmitter and Receiver Optimization for Improper-Complex Data Symbol Transmission over Proper-Complex Cyclostationary Noise}
\title{An Optimal Transmission and Reception of Improper-Complex Data Symbols over Proper-Complex Cyclostationary Noise}
\title{$\;\;$ \\$\;\;$ \\ Joint Transmitter and Receiver Optimization for Improper-Complex Second-Order Stationary Data Sequence}
\vspace{5mm}
\author{Jeongho~Yeo, Joon~Ho~Cho$^\dagger$, and James~S.~Lehnert
\thanks{J.~Yeo and J.~S.~Lehnert are with the School of Electrical and Computer Engineering, Purdue University, West Lafayette, IN 47907-2035 (e-mail: \{jeonghoyeo, lehnert\}@purdue.edu).}
\thanks{J.~H.~Cho is with the Department of Electrical Engineering, Pohang University of Science and Technology (POSTECH), Pohang, Gyeongbuk 790-784, Korea (e-mail: jcho@postech.ac.kr).}
\thanks{$^\dagger$ corresponding author}
}%

\date{ }

\maketitle
\vspace{5mm}
\begin{abstract}
In this paper, the transmission of an improper-complex second-order stationary data sequence is considered over a strictly band-limited frequency-selective channel.
It is assumed that the transmitter employs linear modulation and that the channel output is corrupted by additive proper-complex cyclostationary noise.
Under the average transmit power constraint, the problem of minimizing the mean-squared error at the output of a widely linear receiver is formulated in the time domain to find the optimal transmit and receive waveforms.
The optimization problem is converted into a frequency-domain problem by using the vectorized Fourier transform technique and put into the form of a double minimization.
First, the widely linear receiver is optimized that requires, unlike the linear receiver design with only one waveform, the design of two receive waveforms.
Then, the optimal transmit waveform for the linear modulator is derived by introducing the notion of the impropriety frequency function of a discrete-time random process and by performing a line search combined with an iterative algorithm.
The optimal solution shows that both the periodic spectral correlation due to the cyclostationarity and the symmetric spectral correlation about the origin due to the impropriety are well exploited.
\end{abstract}
\vspace{5mm}
\vspace{10mm}
\section{Introduction}\label{sec: intro}

An information-bearing signal encountered in communications and signal processing often exhibits periodicity in its mean and auto-covariance functions and thus it is well modeled by a wide-sense cyclostationary (WSCS) random process \cite{Proakis_01}.
This structure in the first-order and the second-order statistics has long been exploited in the design of many communications and signal processing systems \cite{Giannakis_05, Gardner_06}.

One of the classical problems related to the processing of WSCS random processes is a joint optimization of the transmitter (Tx) and receiver (Rx) in a communication system.
In \cite{Berger_67, Hansler_71, Ericson_73, Ericson_74}, real-baseband pulse amplitude modulation (PAM) of a wide-sense stationary (WSS) real-valued data symbol sequence is considered with a linear Rx for use over an additive WSS colored noise channel.
Under the minimum mean-squared error (MMSE) optimality criterion and the average transmit power constraint, the jointly optimal transmit and receive waveforms are derived.
It is shown that, interestingly, the waveforms have nonzero spectral values only on a generalized Nyquist interval \cite{Ericson_73} with length equal to the minimum bandwidth required to satisfy the Nyquist condition for zero intersymbol interference (ISI) \cite{Proakis_01}.

This joint optimization problem is extended in \cite{Cho_04} to complex-baseband quadrature amplitude modulation (QAM) of a WSS complex-valued data symbol sequence.
Under the linear MMSE (LMMSE) optimality criterion and the average transmit power constraint, the jointly optimal transmit and receive waveforms are derived for use over an additive WSCS noise channel.
It is well known that a WSCS noise model is better than a WSS model for the case in which data-like QAM interferences are present as well as an ambient Gaussian noise \cite{Proakis_01}.
In contrast to the previous results only with an additive WSS noise, the optimal waveforms are shown in general to have nonzero spectral values on a frequency interval whose length is greater than that of the generalized Nyquist interval.
This is because, unlike a WSS random process, a WSCS random process possesses non-zero correlation in the frequency domain among the components that are spaced integer multiples of the symbol rate apart \cite{Gardner_75}.
To exploit such spectral correlation of the WSCS random process, a vectorized Fourier transform (VFT) technique is employed in \cite{Cho_04}.
This technique is motivated by the harmonic series representation \cite{Gardner_75} of a WSCS random process, and the use of that representation for joint Tx and Rx optimizations in cyclostationary interference and noise has been examined in \cite{Yang_94} and \cite{Golden_95}.

The results in \cite{Yang_94, Golden_95, Cho_04}, however, have considered only the real passband or, equivalently, the complex baseband transmission of a proper-complex data sequence.
Hence, these results are not directly applicable to, e.g., the real passband transmission of a BPSK data sequence, which is an improper-complex data sequence in complex baseband.
Recall that complex-valued random variables, vectors, and processes are called proper if their complementary covariance, complementary covariance matrix, and complementary auto-covariance function (a.k.a.~pseudo-covariance, pseudo-covariance matrix, and pseudo-covariance function) vanish, respectively \cite{Massey_93}.
Otherwise, they are called improper \cite{Schreier_10}.
Although the complex envelopes of the majority of digitally modulated signals are proper, there still remain other digitally modulated signals whose complex envelopes have non-vanishing complementary auto-covariance functions \cite{Schreier_10}.
For example, the complex envelopes of PAM, vestigial sideband PAM, unbalanced QAM, offset quaternary phase-shift keying (OQPSK), and Gaussian minimum shift keying are improper.

Among these improper-complex signals, we focus in this paper on a linear modulation of an improper-complex data sequence using only one transmit waveform.
In particular, we consider an improper-complex data sequence that is well modeled by a zero-mean improper-complex second-order stationary (SOS) random process for which the auto-covariance and the complementary auto-covariance functions depend only on the time difference \cite{Schreier_10}.
This results in an improper-complex second-order cyclostationary (SOCS) transmitted signal.
For example, PAM, vestigial sideband PAM, and unbalanced QAM fall into this category.
It is assumed that such an improper-complex SOCS signal is transmitted over a strictly band-limited frequency-selective linear time-invariant (LTI) channel whose output is corrupted by an additive proper-complex SOCS random process.
As already mentioned, proper-complex SOCS random processes well model the  complex envelopes of the majority of digitally modulated signals as well as the complex envelope of an additive Gaussian noise.

Our objective is to extend the aforementioned joint optimizations of the Tx and Rx for proper-complex WSCS signaling to a joint Tx and Rx optimization problem for improper-complex SOCS signaling under the MMSE optimality criterion and the average transmit power constraint.
It is well known that the second-order properties of an improper-complex signal are not well captured by a linear Rx, but instead by a class of nonlinear Rx's called widely linear Rx's \cite{Schreier_10}.
There are two types of widely linear Rx's.
The first one linearly processes the signal augmented by its complex conjugate, whereas the second one linearly processes the real part of the signal augmented by the imaginary part.
In this paper, the first type of widely linear processing also referred to as the linear-conjugate linear (LCL) filtering \cite{Brown_69} is employed.
It is noteworthy that, unlike the joint optimizations in \cite{Yang_94, Golden_95, Cho_04}, we now need to find two receive waveforms under the widely linear MMSE (WLMMSE) optimality criterion, where one is employed to filter the complex envelope of the received signal and the other to filter its complex conjugate.

The VFT technique again enables us to convert the objective function and the average transmit power constraint described initially in the time domain into those in the frequency domain.
Unlike the previous joint optimizations, the objective function is now expressed in terms of the VFT of the transmit waveform augmented by the VFT of its complex conjugate and the VFT of a receive waveform augmented by the VFT of the other receive waveform.
Using these augmented vector-valued functions, we derive the optimal waveforms of the WLMMSE Rx in a straightforward way as a function of the transmit waveform.
It is shown that the two receive waveforms of the WLMMSE Rx exploit not only the periodic spectral correlation due to the cyclostationarity, but also the symmetric spectral correlation about the origin due to the impropriety \cite{Schreier_10}.

To derive the optimal transmit waveform, we devise the notion of the impropriety frequency function of the transmitted improper-complex SOS data sequence by using the relation between the power spectral density (PSD) and the complementary PSD of the random process.
This real-valued non-negative function converts the transmit waveform optimization problem into an equivalent convex optimization problem to find the optimal energy density of the transmit waveform.
Then, a line search combined with an iterative algorithm is proposed to solve the problem.
After finding the optimal energy density, the optimal transmit and receive waveforms are obtained.
Numerical results provide an example of joint waveform design and also show the effect of the impropriety frequency function on the mean-squared error (MSE) performance.

The rest of this paper is organized as follows.
In Section II, the system model is described and the problem is formulated in the time domain.
In Section III, the problem is reformulated in the frequency domain.
In Section IV, the impropriety frequency function is introduced and the jointly optimal transmit and receive waveforms are derived.
Numerical results are provided in Section V, and concluding remarks are offered in Section VI.

\vspace{-2mm}
\section{System Model and Problem Formulation}\label{sec: signal model}

In this section, we describe the system model and formulate the optimization problem in the time domain.
The system model is an extension of that in \cite{Cho_04}, which only considers the transmission and reception of a proper-complex SOS data sequence, to now allow improper-complex SOS sequences.
The optimality criterion of the joint optimization problem is also extended from the LMMSE criterion to the WLMMSE criterion.

\subsection{System Model}

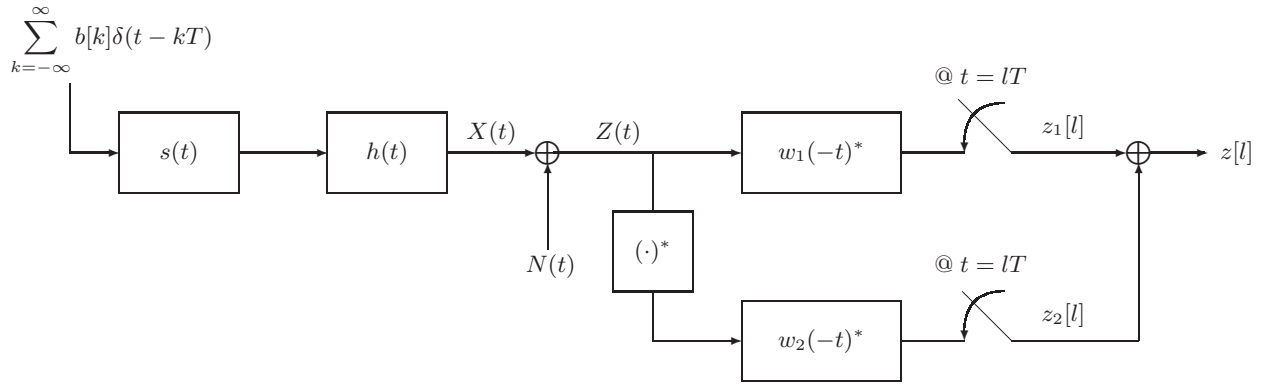
\begin{figure}[tbp]
\setlength{\unitlength}{0.75pt}
{\footnotesize
\begin{center}
\begin{picture}(615,210)(-38,100)

    \put(-20,200){\line(1,0){22}}
    \put(5,200){\vector(1,0){0}}
    \put(-20,235){\line(0,-1){35}}
    \put(-50,255){$\displaystyle\sum_{k=-\infty}^{\infty}b[k]\delta(t-kT)$}

    \put(5,180){\framebox(60,40){$s(t)$}}
    \put(65,200){\line(1,0){40}}
    \put(110,200){\vector(1,0){0}}
    \put(110,180){\framebox(60,40){$h(t)$}}
    \put(180,207){$X(t)$}

    \put(170,200){\line(1,0){40}}
    \put(215,200){\vector(1,0){0}}
    \put(214,197){$\bigoplus$}
    \put(221,151){\line(0,1){40}}
    \put(221,195){\vector(0,1){0}}
    \put(210,140){$N(t)$}
    \put(224,200){\line(1,0){90}}
    \put(319,200){\vector(1,0){0}}
    \put(319,180){\framebox(80,40){$w_1(-t)^*$}}

    \put(245,207){$Z(t)$}
    \put(274,200){\line(0,-1){30}}
    \put(254,130){\framebox(40,40){$(\cdot)^*$}}
    \put(274,130){\line(0,-1){25}}
    \put(274,105){\line(1,0){40}}
    \put(319,105){\vector(1,0){0}}
    \put(319,85){\framebox(80,40){$w_2(-t)^*$}}

    \put(431,200){
    	\put(-32,0){\line(1,0){32}}
    	{\bezier{500}(0,04)(0,25)(20,25)}
	\put(24,0){\line(-1,1){27}}
    	\vector(0,-1){0}
	\put(-15,35){$@\; t=lT$}}

    \put(455,200){\line(1,0){53}}
    \put(513,200){\vector(1,0){0}}
    \put(470,210){$z_1[l]$}
    \put(512,197){$\bigoplus$}
    \put(525,200){\line(1,0){27}}
    \put(554,200){\vector(1,0){0}}
    \put(560,197){$z[l]$}

    \put(431,105){
    	\put(-32,0){\line(1,0){32}}
    	\bezier{500}(0,04)(0,25)(20,25)
	\put(24,0){\line(-1,1){25}}
    	\vector(0,-1){0}
	\put(-15,35){$@\; t=lT$}}

    \put(455,105){\line(1,0){64}}
    \put(470,115){$z_2[l]$}
    \put(519,105){\line(0,1){85}}
    \put(519,195){\vector(0,1){0}}

\end{picture}
\end{center}
}%
\caption{System block diagram.} \label{Fig: System block diagram}
\end{figure}

A Tx and an Rx operate over a real passband to transmit a data sequence $\{b[l]\}_{l\in\mathbb{Z}}$.
Fig.~\ref{Fig: System block diagram} shows the system block diagram in complex baseband.
The data sequence $\{b[l]\}_{l\in\mathbb{Z}}$ is assumed well modeled by a zero-mean improper-complex SOS random process with auto-covariance and complementary auto-covariance functions given, respectively, by $m[k] \triangleq  {\mathbb{E}}\{b[k+l]b[l]^*\}$ and $\tilde{m}[k] \triangleq  {\mathbb{E}}\{b[k+l]b[l]\}$,
where the superscript $^*$ denotes complex conjugation.
By applying the discrete-time Fourier transform (DTFT) operations to $m[k]$ and $\tilde{m}[k]$, the PSD $M(f)$ and the complementary PSD $\tilde{M}(f)$ of the data sequence $\{b[l]\}_{l\in\mathbb{Z}}$ are derived, respectively, as
$M(f) \triangleq  \sum_{k=-\infty}^{\infty} m[k] e^{-{\rm j} 2 \pi fk}$ and $\tilde{M}(f) \triangleq  \sum_{k=-\infty}^{\infty} \tilde{m}[k] e^{-{\rm j} 2 \pi fk}$.

The Tx to be designed employs linear modulation with symbol transmission rate $1/T$ [symbols/sec], where the transmit waveform is denoted by $s(t)$.
The transmitted signal $\sum_{k=-\infty}^{\infty} b[k] s(t - kT)$ is passed through a strictly band-limited channel that is modeled by an LTI system with impulse response $h(t)$ having the one-sided bandwidth $B$ [Hz] in complex baseband.

The received signal denoted by $Z(t)$ consists of the signal from the Tx and an additive interference-plus-noise signal $N(t)$, where the latter is modeled by a zero-mean proper-complex SOCS random process with fundamental cycle period $T_0$.
It is assumed that the multiplicative inverse $T$ of the symbol transmission rate of the desired signal is chosen as an integer multiple of $T_0$.
Thus, $Z(t)$ can be written as
\vspace{-6mm}
\begin{equation}\label{eq: signal_model}
Z(t) = \sum_{k=-\infty}^{\infty} b[k] p(t - kT) + N(t),\vspace{-4mm}
\end{equation}
where $p(t)\triangleq h(t)*s(t)$ denotes the overall response with the operator $*$ denoting the convolution integral.
There should be no confusion from the superscript $^*$ that denotes the complex conjugation.

In (\ref{eq: signal_model}), it can be easily shown that the desired signal component $X(t)\triangleq \sum_{k=-\infty}^{\infty} b[k] p(t - kT)$ becomes a zero-mean SOCS random process due to the second-order property of the zero-mean SOS data sequence $\{b[l]\}_{l\in\mathbb{Z}}$.
In other words, the mean, the auto-covariance, and the complementary auto-covariance functions of $X(t)$ satisfy, respectively, $\mu_X(t) \triangleq {\mathbb{E}}\{X(t)\} = 0$, $r_X(t,s) \triangleq {\mathbb{E}}\{X(t)X(s)^*\} = r_X(t+T,s+T)$, and $\tilde{r}_X(t,s) \triangleq {\mathbb{E}}\{X(t)X(s)\} = \tilde{r}_X(t+T,s+T)$, $\forall t, \forall s$.
In what follows, we also call $r_X(t,s)$ and $\tilde{r}_X(t,s)$ the auto-correlation and the complementary auto-correlation functions, respectively, because $X(t)$ has mean zero.

In (\ref{eq: signal_model}), it can be straightforwardly shown that the interference-plus-noise signal $N(t)$ is SOCS with mean zero and cycle period $T$, because $T$ is assumed to be an integer multiple of $T_0$, i.e., $\mu_N(t) \triangleq {\mathbb{E}}\{N(t)\} = 0$, $r_N(t,s) \triangleq {\mathbb{E}}\{N(t)N(s)^*\} = r_N(t+T,s+T)$, and $\tilde{r}_N(t,s) \triangleq {\mathbb{E}}\{N(t)N(s)\} = \tilde{r}_N(t+T,s+T)$, $\forall t, \forall s$.
Now that $Z(t)$ is a summation of two uncorrelated zero-mean SOCS random processes with cycle period $T$, it is also a zero-mean SOCS random processes with cycle period $T$.
%


It is well known \cite{Schreier_10} that, for a vector-valued signal model, a widely linear Rx employing two linear filters outperforms a linear Rx employing only one linear filter when either the desired signal or the interference-plus-noise signal is improper.
Thus, in this paper, we employ two LTI filters with impulse responses $w_1(-t)^*$ and $w_2(-t)^*$ to process the improper-complex SOCS process $Z(t)$ and its complex conjugate $Z(t)^*$, respectively.
The two LTI filters are followed by uniform samplers with rate $1/T$ [samples/sec], and then the sequence of decision statistics $\{z[l]\}_{l\in\mathbb{Z}}$ is obtained as the sum of the sampler outputs, i.e.,
\vspace{-7mm}
\begin{equation}
z[l] \triangleq z_{1}[l] + z_{2}[l],\vspace{-7mm}
\end{equation}
where the sampler outputs $z_{1}[l]$ and $z_{2}[l]$ are defined, respectively, as
\vspace{-7mm}
\begin{IEEEeqnarray}{rCl}
z_{1}[l] &\triangleq & w_1(-t)^* * Z(t) \big|_{t=lT} = \! \int_{-\infty}^{\infty} w_1(t-lT)^* Z(t) dt \quad \text{and} \IEEEeqnarraynumspace\IEEEyesnumber\IEEEyessubnumber\\
z_{2}[l] &\triangleq & w_2(-t)^* * Z(t)^* \big|_{t=lT} \! = \! \! \int_{-\infty}^{\infty} \! w_2(t-lT)^* Z(t)^* dt. \;\;\IEEEeqnarraynumspace\IEEEyessubnumber\vspace{-7mm}
\end{IEEEeqnarray}

\subsection{Problem Formulation in Time Domain}
Our objective is to find the transmit and receive waveforms $s(t)$, $w_1(t)$, and $w_2(t)$ that jointly minimize the MSE given by
\vspace{-4mm}
\begin{equation}\label{eq: MSE_TD}
\varepsilon \big( s(t), w_1(t), w_2(t) \big) \triangleq {\mathbb{E}}\{|b[l] - z[l] |^2\},\vspace{-4mm}
\end{equation}
where $s(t)$, $w_1(t)$, and $w_2(t)$ are the parameters to be designed.
Since $T$ is an integer multiple of the fundamental cycle period $T_0$ of the interference-plus-noise signal, it can be easily shown that the MSE defined in (\ref{eq: MSE_TD}) as the objective function of the optimization problem is the same regardless of the value of $l$.

The average transmit power constraint is then imposed on this joint optimization problem.
Since the transmitted signal is SOCS with cycle period $T$, the average transmit power $\bar{P}$ can be defined as
\vspace{-4mm}
\begin{equation}\label{eq: power_constraint_TD}
\bar{P} \triangleq {\mathbb{E}}\left\{ \frac{1}{T}\int_{\langle T \rangle}  \left|\sum_{k=-\infty}^{\infty} b[k] s(t - kT) \right|^2dt \right\},\vspace{-4mm}
\end{equation}
where $\langle T \rangle$ denotes any integration interval of length $T$ [sec].
Thus, the constraint is given by $\bar{P}=P_T$ for some $P_T>0$.
Therefore, the joint optimization problem is given by
\begin{problem}\label{problem: 1}
\end{problem}
\vspace{-27mm}
\begin{IEEEeqnarray}{ll}
\underset{s(t),\; w_1(t),\; w_2(t)}{\text{minimize}}& \quad \varepsilon \big( s(t), w_1(t), w_2(t) \big)\IEEEeqnarraynumspace\IEEEyesnumber\IEEEyessubnumber\label{eq: problem}\\
\;\;\,\text{subject to}\;\;& \quad \bar{P} = P_T. \IEEEeqnarraynumspace\IEEEyessubnumber \label{eq: power constraint}\vspace{-5mm}
\end{IEEEeqnarray}

\section{Problem Reformulation in Frequency Domain}\label{sec: MSE_FD}
In this section, Problem~\ref{problem: 1} described in the time domain is reformulated in the frequency domain.
To proceed, we first review the notions of the VFT and the matrix-valued PSD.
Then, by proposing the notion of the matrix-valued complementary PSD and the methods to augment the VFTs of the transmit and receive waveforms, we convert the objective function (\ref{eq: MSE_TD}) and the average transmit power constraint (\ref{eq: power_constraint_TD}) to equivalent expressions in the frequency domain.

\subsection{Review of VFT and Matrix-Valued PSD}\label{section: subsection-review}

In this subsection, we briefly review the notions of excess bandwidth, the Nyquist interval, the VFT, and the matrix-valued PSD.
For details, see \cite{Cho_04}.

Given a pair $(B,1/T)$ of a bandwidth and a reference rate, the excess bandwidth $\beta$ is defined as $\beta \triangleq 2BT - 1$ and the Nyquist interval ${\mathcal{F}}$ is defined as ${\mathcal{F}} \triangleq \left\{f: -\frac{1}{2T} \leq f < \frac{1}{2T} \right\}$.

Given a pair $(B,1/T)$ and a deterministic function $p(t)$ having the continuous-time Fourier transform (CTFT) $P(\xi) \triangleq \int_{-\infty}^{\infty} p(t) e^{-{\rm j}2\pi \xi t}dt$, the VFT ${\bm{p}}(f)$ of $p(t)$ is defined as a vector-valued function of $f\in{\mathcal{F}}$ that is equivalent to $P(\xi)$.
In particular, the $k$th entry of ${\bm{p}}(f)$ is given by $[{\bm{p}}(f)]_k\triangleq P\left(f+\frac{k-L-1}{T}\right)$
for $k=1,2,\cdots,2L+1$, where $L\triangleq \lceil \beta/2 \rceil$.

Given a pair $(B,1/T)$ and an SOCS random process $N(t)$ with cycle period $T$ having the auto-correlation function $r_N(t,s)$, the matrix-valued PSD ${\bm{R}}_N(f)$ of $N(t)$ is defined as a matrix-valued function of $f\in{\mathcal{F}}$, whose $(k,l)$th entry is given by $
[{\bm{R}}_N(f)]_{k,l}\triangleq R_N^{(k-l)} (f+({l-L-1})/{T})$
for $k, l = 1,2,\cdots,2L+1$, where $R_N^{(k)}(\xi)$ is the CTFT of $r_N^{(k)}(\tau)$ that is obtained by applying the Fourier series expansion to $r_N(t,t-\tau)$, i.e., $r_N(t,s)=\sum_{k=-\infty}^{\infty}r_N^{(k)}(t-s)e^{{\rm j} 2\pi k t/T}$.
%

In using the above definitions, it is assumed that the parameter $B$ is chosen as bandwidth in complex baseband over which the Rx can observe and process a signal and that the parameter $1/T$ is chosen as the symbol transmission rate of the Tx.
It is also assumed that the frequency band over which the Tx can emit non-zero power is identical to the frequency band of the Rx.
For a general case where these two frequency bands are different, the notion of virtual legacy Rx's and the orthogonal constraint at the virtual legacy Rx's can be employed as is done in \cite{Yun_10} for the transmission of a proper-complex data sequence.

Due to the above assumption on the frequency band that can be used by the Tx and the Rx, the first and the last entries of the VFT of the transmit waveform need to be always zero for $-1/(2T) \leq f \leq L/T - B$ and $ B - L/T \leq f \leq 1/(2T) $, respectively.
For this, the notion of the effective VFT is employed as discussed in \cite{Cho_04, Yun_10}, and \cite{Cho_05}.
The effective VFT is defined as a variable-length vector-valued function of $f\in{\mathcal{F}}$ by removing the first and the last entries of the VFT for $-1/(2T) \leq f \leq L/T - B$ and $ B - L/T \leq f \leq 1/(2T) $, respectively.
In what follows, the length of the effective VFT is denoted by ${\mathcal{N}}(f)$.
For details, see \cite[Eq.~(14)]{Cho_05}.
Similarly, the effective matrix-valued PSD can be also defined as an ${\mathcal{N}}(f)$-by-${\mathcal{N}}(f)$ matrix-valued function of $f\in{\mathcal{F}}$ by removing both the first row and column of the matrix-valued PSD for $-1/(2T) \leq f \leq L/T - B$ and by removing both the last row and column for $ B - L/T \leq f \leq 1/(2T) $.

\subsection{Problem Reformulation in Frequency Domain}

In this subsection, the objective function and the average transmit power constraint in Problem~\ref{problem: 1} are converted into equivalent expressions in the frequency domain.
To begin with, we propose the notion of the matrix-valued complementary PSD of an improper-complex SOCS random process.
\begin{definition}\label{def: com_mv_psd}
Given a pair $(B,1/T)$ and an improper-complex SOCS random process $X(t)$ with cycle period $T$ and complementary auto-correlation function $\tilde{r}_X(t,s)$, let $\tilde{R}_X^{(k)}(\xi)$ be the CTFT of $\tilde{r}_X^{(k)}(\tau)$ that is obtained by applying the Fourier series expansion to the periodic signal $\tilde{r}_X(t,t-\tau)=\tilde{r}_X(t+T,t+T-\tau)$, $\forall t$, i.e., $\tilde{r}_X(t,s)=\sum_{k=-\infty}^{\infty}\tilde{r}_X^{(k)}(t-s)e^{{\rm j} 2\pi k t /T}$.
Then, the matrix-valued complementary PSD $\tilde{\bm{R}}_X(f)$ is defined as a matrix-valued function of $f\in{\mathcal{F}}$, whose $(k,l)$th entry is given by $[\tilde{\bm{R}}_X(f)]_{k,l}\triangleq \tilde{R}_X^{(k-l)}\left(f+({l-L-1})/{T}\right)$
for $k,l=1,\cdots,2L+1$.
\end{definition}


Note that the matrix-valued complementary PSD $\tilde{\bm{R}}_N(f)$ of the interference-plus-noise signal $N(t)$ becomes an all-zero matrix because $N(t)$ is modeled by a zero-mean proper-complex SOCS random process.
Note also that the effective matrix-valued complementary PSD can be defined similarly to the effective matrix-valued PSD.
In what follows, each of the VFT, the matrix-valued PSD, and the matrix-valued complementary PSD is an effective one.

By using the above definitions, the matrix-valued PSD and the matrix-valued complementary PSD of the desired signal component in (\ref{eq: signal_model}) are derived as follows.

\begin{lemma}\label{lemma: MV-PSD_2}
The ${\mathcal{N}}(f)$-by-${\mathcal{N}}(f)$ matrix-valued PSD $\bm{R}_X(f)$ and the ${\mathcal{N}}(f)$-by-${\mathcal{N}}(-f)$ matrix-valued complementary PSD $\tilde{\bm{R}}_X(f)$ of the desired signal $X(t)= \sum_{l=-\infty}^{\infty} b[l] p(t - lT)$ are given by
\vspace{-5mm}
\begin{IEEEeqnarray}{C}
\bm{R}_X(f) = \frac{1}{T} M(fT) \bm{p}(f)\bm{p}(f)^{\mathcal{H}} \quad \text{and} \quad \tilde{\bm{R}}_X(f) = \frac{1}{T} \tilde{M}(fT) \bm{p}(f)\big( \bm{J}(-f) \bm{p}(-f)^*  \big)^{\mathcal{H}},  \IEEEeqnarraynumspace\IEEEyesnumber\label{eq: com_MV-PSD-LM}\vspace{-5mm}
\end{IEEEeqnarray}
respectively, where $\bm{p}(f)$ denotes the VFT of $p(t)$, $\bm{J}(f)$ denotes the ${\mathcal{N}}(f)$-by-${\mathcal{N}}(f)$ backward identity matrix whose $(m,n)$th entry is given by $1$ for $m+n={\mathcal{N}}(f)+1$, and $0$ otherwise, and $^{\mathcal{H}}$ denotes Hermitian transposition.
\end{lemma}

\begin{IEEEproof}
By using the CTFT of $r_X^{(k)}(\tau)$ and $\tilde{r}_X^{(k)}(\tau)$, it can be easily shown that $R_X^{(k)}(f)  = M(fT) P (f + k/T)P(f)^* /T$ and $\tilde{R}_X^{(k)}(f) =  \tilde{M}(fT)P(f+k/T) P(-f)/T$.
Therefore, the conclusion follows from the definitions reviewed in Section~\ref{section: subsection-review} and Definition~\ref{def: com_mv_psd}.
\end{IEEEproof}

Note that $\bm{J}(-f)\bm{p}(-f)^*$ in (\ref{eq: com_MV-PSD-LM}) is nothing but the VFT of $p(t)^*$.
Thus, $\tilde{\bm{R}}_X(f)$ can be interpreted as the correlation between the frequency components at $f$ of $X(t)$ and $X(t)^*$.

%

Now, we are ready to convert the objective function.
The MSE $\varepsilon\triangleq \varepsilon \big( s(t), w_1(t), w_2(t) \big)$ defined in (\ref{eq: MSE_TD}) can be rewritten as
\vspace{-6mm}
\begin{equation}\label{eq: MSE WL}
\varepsilon = {\mathbb{E}}\{|b[l] |^2\} -  2 \Re \Big({\mathbb{E}}\{b[l] ^* z_1[l]\}\Big)  + {\mathbb{E}}\{|z_1[l] |^2\} -  2 \Re \Big({\mathbb{E}}\{b[l] ^* z_2[l] \}\Big)  +  2\Re\Big( {\mathbb{E}}\{z_1[l] z_2[l]^*  \} \Big) + {\mathbb{E}}\{| z_2[l] |^2\},   \vspace{-6mm}
\end{equation}
where $\Re (\cdot)$ denotes the real part.
In the following propositions, each component of the right side of (\ref{eq: MSE WL}) is expressed in terms of the VFT, the matrix-valued PSD, and the matrix-valued complementary PSD.

\begin{proposition}\label{proposition: proof1}
The first three terms of the right side of (\ref{eq: MSE WL}) can be rewritten as ${\mathbb{E}}\{|b[l] |^2\} =  \int_{{\mathcal{F}}} TM(fT) df$, ${\mathbb{E}}\{b[l] ^* z_1[l]\} = \int_{{\mathcal{F}}}   \bm{w}_1(f)^{\mathcal{H}} M(fT) \bm{p}(f) df$, and ${\mathbb{E}}\{|z_1[l] |^2\} = \int_{{\mathcal{F}}}   \bm{w}_1(f)^{\mathcal{H}}   \bm{R}(f) \bm{w}_1(f)df$, respectively, where $\bm{w}_1(f)$ is the VFT of $w_1(t)$ and $\bm{R}(f) \triangleq \bm{R}_N(f) + \bm{R}_X(f)$.
\end{proposition}


\begin{IEEEproof}
See \cite[Proposition~1-4]{Cho_04}.
\end{IEEEproof}

\begin{proposition}\label{proposition: proof2}.
The last three terms of the right side of (\ref{eq: MSE WL}) can be rewritten as ${\mathbb{E}}\{b[l] ^* z_2[l]\} = \int_{{\mathcal{F}}}  \bm{w}_2(f)^{\mathcal{H}}$ $ \tilde{M}(fT)^* \bm{J}(-f) \bm{p}(-f)^* df$, ${\mathbb{E}}\{z_1[l] z_2[l]^*\} = \int_{{\mathcal{F}}}  \bm{w}_1(f)^{\mathcal{H}} \tilde{\bm{R}}(f) \bm{w}_2(f) df$, and ${\mathbb{E}}\{|z_2[l] |^2\} = \int_{{\mathcal{F}}}  \bm{w}_2(f)^{\mathcal{H}}  \bm{J}(-f) $ $\bm{R}(-f)^* \bm{J}(-f) \bm{w}_2(f) df$, respectively, where $\bm{w}_2(f)$ is the VFT of $w_2(t)$ and $\tilde{\bm{R}}(f) \triangleq \tilde{\bm{R}}_X(f)$.
\end{proposition}


\begin{IEEEproof}
It can be shown similarly to Proposition~\ref{proposition: proof1}.
\end{IEEEproof}

Note in ${\mathbb{E}}\{|z_2[l] |^2\} = \int_{{\mathcal{F}}}  \bm{w}_2(f)^{\mathcal{H}}  \bm{J}(-f) $ $\bm{R}(-f)^* \bm{J}(-f) \bm{w}_2(f) df$ that the pre-multiplication of the backward identity matrix $\bm{J}(f)$ reverses the order of the rows whereas the post-multiplication reverses that of the columns.
Note also that $\bm{p}(f) = \bm{H}(f)\bm{s}(f), \; \forall f\in{\mathcal{F}}$, where $\bm{s}(f)$ is the VFT of the transmit waveform $s(t)$ and $\bm{H}(f)$ is defined as $\bm{H}(f) \triangleq \text{diag} \big\{ \bm{h}(f) \big\}$ with $\bm{h}(f)$ representing the VFT of $h(t)$.

To simplify the expression of the objective function, we define $\bar{\mathcal{N}}(f)\triangleq {\mathcal{N}}(f) + {\mathcal{N}}(-f)$,
\vspace{-7mm}
\begin{IEEEeqnarray}{C}\label{eq: def_augmented_VFT}
\bar{\bm{s}}(f) \triangleq \big[ \bm{s}(f)^{\mathcal{T}},\;  \big( \bm{J}(-f) \bm{s}(-f)^* \big)^{\mathcal{T}} \big]^{\mathcal{T}} \quad \text{and} \quad \bar{\bm{w}}(f) \triangleq  \big[ \bm{w}_1(f)^{\mathcal{T}},\; \bm{w}_2(f)^{\mathcal{T}} \big]^{\mathcal{T}},\IEEEeqnarraynumspace\IEEEyesnumber\vspace{-7mm}
\end{IEEEeqnarray}
where $^{\mathcal{T}}$ denotes transposition.
Here, the length-$\bar{\mathcal{N}}(f)$ vector-valued functions $\bar{\bm{s}}(f)$ and $\bar{\bm{w}}(f)$ are the VFT of the transmit waveform augmented by the VFT of its complex conjugate and the VFT of a receive waveform augmented by the VFT of the other receive waveform, respectively.
Also, let the $\bar{\mathcal{N}}(f)$-by-$\bar{\mathcal{N}}(f)$ matrices $\bar{\bm{H}}(f)$, $\bar{\bm{M}}(f)$, and $\bar{\bm{R}}(f)$ be defined, respectively, as $\bar{\bm{H}}(f)  \triangleq \text{diag}\big\{ \bm{H}(f), $ $\bm{J}(-f) \bm{H}(-f)^* \bm{J}(-f) \big\}$, $\bar{\bm{M}}(f)  \triangleq \text{diag}\big\{ M(f) \bm{I}(f), \tilde{M}(f)^*$ $\bm{I}(-f) \big\}$, and
\vspace{-5mm}
\begin{equation}\label{eq: aug_PSD}
\bar{\bm{R}}(f) \triangleq
\left[
\begin{array}{cc}
\,\vspace{-0.75in}\\
\bm{R}(f) & \tilde{\bm{R}}(f)\\
\,\vspace{-0.80in}\\
\tilde{\bm{R}}(f)^{\mathcal{H}} & \bm{J}(-f) \bm{R}(-f)^* \bm{J}(-f)\,\vspace{-0.10in}\\
\end{array}
\right] \vspace{-4mm}
\end{equation}
with $\bm{I}(f)$ denoting the ${\mathcal{N}}(f)$-by-${\mathcal{N}}(f)$ identity matrix and $\text{diag}\{\bm{A}, \bm{B}\}$ denoting the block diagonal matrix whose diagonal blocks are the matrices $\bm{A}$ and $\bm{B}$.
These notions enable us to derive the optimal receive waveforms in a straightforward way.

By substituting the results of Propositions~\ref{proposition: proof1} and \ref{proposition: proof2} into (\ref{eq: MSE WL}), we can rewrite the objective function $\varepsilon$ as
\vspace{-7mm}
\begin{equation}\label{eq: MSE_freq}
\varepsilon \big(\bar{\bm{s}}(f), \bar{\bm{w}}(f) \big)  = \int_{{\mathcal{F}}}  \Big(TM(fT) + \bar{\bm{w}}(f)^{\mathcal{H}} \bar{\bm{R}}(f) \bar{\bm{w}}(f) - 2 \Re \{\bar{\bm{w}}(f)^{\mathcal{H}} \bar{\bm{H}}(f)\bar{\bm{M}}(fT) \bar{\bm{s}}(f)   \} \Big) df,  \vspace{-7mm}
\end{equation}
which is a function of $\bar{\bm{s}}(f)$ and $\bar{\bm{w}}(f)$.
Also, by using \cite[Eq.~(32)]{Cho_04} and the definition of $\bar{\bm{s}}(f)$, we can rewrite the average transmit power $\bar{P}$ defined in (\ref{eq: power_constraint_TD}) as
\vspace{-5mm}
\begin{equation}\label{eq: constraint_FD_b}
\bar{P} = \frac{1}{T}\int_{{\mathcal{F}}} M(fT) \bm{s}(f)^{\mathcal{H}} \bm{s}(f)df = \frac{1}{2T}\int_{{\mathcal{F}}} M(fT) \bar{\bm{s}}(f) ^{\mathcal{H}} \bar{\bm{s}}(f) df.\vspace{-5mm}
\end{equation}

This leads to the equivalent joint optimization problem to find $\bar{\bm{s}}(f)$ and $\bar{\bm{w}}(f)$ as
\begin{problem}\label{problem: FD}
\end{problem}
\vspace{-27mm}
\begin{IEEEeqnarray}{ll}
\underset{\bar{\bm{s}}(f),\; \bar{\bm{w}}(f)}{\text{minimize}}&  \varepsilon \big(\bar{\bm{s}}(f), \bar{\bm{w}}(f) \big)  \IEEEeqnarraynumspace\IEEEyesnumber\IEEEyessubnumber\label{eq: problem_MSE_FD}\\
\,\text{subject to}\;\;& \bar{P} = P_T.  \IEEEeqnarraynumspace\IEEEyessubnumber\label{eq: problem_constraint_FD} \vspace{-7mm}
\end{IEEEeqnarray}
In the next section, we solve this optimization problem to obtain the VFTs of the optimal receive and transmit waveforms.

\section{Optimization of Transmit and Receive Waveforms}\label{sec: optimal_solution}

In this section, we first derive the optimal $\bar{\bm{w}}(f)$ that minimizes the objective function in (\ref{eq: problem_MSE_FD}) for a given $\bar{\bm{s}}(f)$.
Then, by substituting this $\bar{\bm{w}}(f)$ and introducing the notion of the impropriety frequency function, we obtain the optimization problem   over ${\bm{s}}(f)$.
By solving this problem, we finally obtain the optimal transmit and receive waveforms.


%
%
%
%

\subsection{Optimization of Widely Linear Receiver }

As in \cite[Theorem~2]{Cho_04}, to find the optimal $\bar{\bm{w}}(f)$ for given $\bar{\bm{s}}(f)$, an unconstrained quadratic optimization problem is solved.
Thus, by using the first-order necessary condition, we have the solution
\vspace{-5mm}
\begin{equation}\label{eq: Rx_opt}
\bar{\bm{w}}(f) =  \bar{\bm{R}}(f)^{-1} \bar{\bm{H}}(f) \bar{\bm{M}}(fT) \bar{\bm{s}}(f) , \; \forall f\in {\mathcal{F}}.\vspace{-7mm}
\end{equation}
By substituting the above solution into (\ref{eq: MSE_freq}), we can rewrite the MSE as
\vspace{-7mm}
\begin{IEEEeqnarray}{l}\label{eq: Rx_opt_MSE}
\varepsilon \big(\bar{\bm{s}}(f)\big)   = \int_{{\mathcal{F}}}  \Big(TM(fT) -  \bar{\bm{s}}(f) ^{\mathcal{H}}  \bar{\bm{M}}(fT)^{\mathcal{H}} \bar{\bm{H}}(f)^{\mathcal{H}}  \bar{\bm{R}}(f)^{-1} \bar{\bm{H}}(f) \bar{\bm{M}}(fT) \bar{\bm{s}}(f)  \Big) df.\IEEEeqnarraynumspace\IEEEyesnumber\vspace{-5mm}
\end{IEEEeqnarray}
which is a function only of $\bar{\bm{s}}(f)$.
%
%

\subsection{Impropriety Frequency Function}

To convert $\varepsilon \big(\bar{\bm{s}}(f)\big)$ into a function only of $\bm{s}(f)$, the notion of the impropriety frequency function is introduced as follows.

\begin{definition}
Given a discrete-time improper-complex SOS random process with PSD $M(f)$ and complementary PSD $\tilde{M}(f)$, its impropriety frequency function $k(f)$ is defined as
\vspace{-4mm}
\begin{equation}\label{eq: def_k}
k(f) \triangleq \left\{
\begin{array}{ll}
\,\vspace{-0.80in}\\
0,& \text{ if } M(f)M(-f)=0, \\
\,\vspace{-0.80in}\\
\displaystyle \frac{|\tilde{M}(f)|}{\sqrt{M(f)M(-f)}},& \text{ otherwise.}
\end{array}
\right. \vspace{3mm}
\end{equation}
\end{definition}
\vspace{1mm}

The above definition is motivated by the impropriety coefficient of an improper-complex random variable \cite[Definition~3.1]{Schreier_10} and by a relation between $M(f)$ and $\tilde{M}(f)$ shown in \cite[Eq.~(5)]{Picinbono_97}.
By using the phase $\phi(f)$ of $\tilde{M}(f)$, we can rewrite the complementary PSD as $\tilde{M}(f) =|\tilde{M}(f)| e^{{\rm j} \phi(f)}= k(f)\sqrt{M(f)M(-f)}e^{{\rm j} \phi(f)}$, where $0 \leq \phi(f) \leq 2\pi$.
In the next lemma, the properties of the impropriety frequency and the phase functions are provided.

\begin{lemma}\label{lemma: PSD_com}
The impropriety frequency function $k(f)$ and the phase function $\phi(f)$ satisfy
\vspace{-7mm}
\begin{IEEEeqnarray}{C}
0 \leq k(f) \leq 1, \quad  k(-f)=k(f), \quad \text{and} \quad  \phi(-f)=\phi(f), \; \forall f.\IEEEeqnarraynumspace\IEEEyesnumber \vspace{0mm}
\end{IEEEeqnarray}
\end{lemma}

\begin{IEEEproof}
Since $\tilde{m}[-k]=\tilde{m}[k]$ by definition, we have $\tilde{M}(-f)=\tilde{M}(f)$, which implies $\phi(-f)=\phi(f)$.
This also leads to $k(-f)=k(f)$ by (\ref{eq: def_k}).
By using the property $| \tilde{M}(f) |^2 \leq M(f)M(-f)$ shown in \cite[Eq.~(5)]{Picinbono_97}, we have $0 \leq k(f) \leq 1$.
\end{IEEEproof}

For example, an uncorrelated real-valued PAM data sequence results in $k(f)=1,\forall f$, whereas any proper-complex data sequence results in $k(f)=0,\forall f$.
By using the impropriety frequency function, we can rewrite the MSE (\ref{eq: Rx_opt_MSE}) in the form of a function of $\bar{\bm{s}}(f)$ as a function of $\bm{s}(f)$.

\begin{lemma}\label{lemma: MSE_Rx}
Define $c(f)$ as
\vspace{-8mm}
\begin{equation}\label{eq: c_f}
c(f)\triangleq \frac{1}{T}M(fT)\bm{s}(f)^{\mathcal{H}}\bm{H}(f)^{\mathcal{H}}\bm{R}_N(f)^{-1}\bm{H}(f)\bm{s}(f). \vspace{-6mm}
\end{equation}
By using $c(f)$ and $k(f)$, also define $\bm{D}(f)$ and $\bm{k}(f)$, respectively, as
\vspace{-5mm}
\begin{equation}\label{eq: c_matrix_f}
\bm{D}(f) \triangleq \begin{bmatrix}
\,\vspace{-0.80in}\\
c(f) &\!\! 0 \\
\,\vspace{-0.80in}\\
0 & \!\! \frac{c(-f)}{1 + c(-f)(1 - k(fT)^2)} \,\vspace{-0.1in}\\
\end{bmatrix} \quad \text{and} \quad \bm{k}(f) \triangleq \begin{bmatrix}
\,\vspace{-0.80in}\\
1  \\
\,\vspace{-0.80in}\\
k(f)  \,\vspace{-0.1in}\\
\end{bmatrix}.\vspace{-4mm}
\end{equation}
Then, the MSE $\varepsilon \big( \bar{\bm{s}}(f) \big)$ in (\ref{eq: Rx_opt_MSE}) can be rewritten as
\vspace{-5mm}
\begin{equation}\label{eq: MSE_simple}
\varepsilon \big(\bm{s}(f)\big) = \int_{{\mathcal{F}}}  \frac{TM(fT)}{1+\bm{k}(fT)^{\mathcal{T}} \bm{D}(f)\bm{k}(fT)  } df,\vspace{-6mm}
\end{equation}
which is a function of $\bm{s}(f)$.
\end{lemma}

\begin{IEEEproof}
See Appendix~B.
\end{IEEEproof}


\subsection{Optimization of Transmitter}

Let $\varepsilon (f)$ denote the integrand in (\ref{eq: MSE_simple}), i.e., $\varepsilon (f) \triangleq TM(fT)/(1+\bm{k}(fT)^{\mathcal{T}} \bm{D}(f)\bm{k}(fT) )$.
Then, by the definitions of $c(f)$ and $\bm{D}(f)$ in (\ref{eq: c_f}) and (\ref{eq: c_matrix_f}), respectively, it can be seen that $\varepsilon (f_0)$ for some $f_0$ is affected by the choice of $\bm{s}(f)$ at both $f_0$ and $-f_0$.
Thus, Problem~\ref{problem: FD} can be rewritten as
\begin{problem}\label{problem: FD_TX}
\end{problem}
\vspace{-20mm}
\begin{IEEEeqnarray}{ll}
\underset{a(f), \; a(-f) }{\text{minimize}}& \left\{\begin{array}{ll}
\,\vspace{-0.80in}\\
\underset{\bm{s}(f), \; \bm{s}(-f) }{\text{minimize}} \displaystyle \;\; \int_{{\mathcal{F}}^+}  \varepsilon (f) + \varepsilon (-f) df \IEEEeqnarraynumspace\IEEEyesnumber\IEEEyessubnumber\label{problem: inner}\\
\,\vspace{-0.70in}\\
\text{subject to} \;\; \displaystyle \| \bm{s}(f) \|^2 = a(f), \;\forall f\in {\mathcal{F}} \,\vspace{-0.05in}\\
\end{array} \right.\IEEEeqnarraynumspace\IEEEyessubnumber\\
\,\text{subject to}\;\;& \frac{1}{T}\int_{{\mathcal{F}}^+} M(fT) a(f) + M(-fT) a(-f)df = P_T, \IEEEeqnarraynumspace\IEEEyessubnumber \vspace{-5mm}
\end{IEEEeqnarray}
where $a(f) \triangleq \| \bm{s}(f) \|^2$ is the energy density of $\bm{s}(f)$ and ${\mathcal{F}^+} \triangleq \{f: 0 \leq f < 1/(2T) \}$ denotes the half-Nyquist interval.
Note that the problem is now in the form of a double minimization problem, where the constraint set of $\bm{s}(f)$ is partitioned into subsets, each of which has all $\bm{s}(f)$ having the same $a(f)$.

\begin{proposition}\label{proposition: inner_optimization}
Given $a(f)$, the optimal solution to the inner optimization problem in (\ref{problem: inner}) is given by
\vspace{-7mm}
\begin{equation}\label{eq: opt_TX_waveform}
\bm{s}(f) = \sqrt{a(f)}\bm{v}(f)e^{{\rm j} \theta(f)}, \;\forall f\in {\mathcal{F}},\vspace{-7mm}
\end{equation}
where $\bm{v}(f)$ is the normalized eigenvector corresponding to the largest eigenvalue of $\bm{H}(f)^{\mathcal{H}}\bm{R}_N(f)^{-1}\bm{H}(f)$, and $\theta(f)$ can be chosen arbitrarily.
\end{proposition}

\begin{IEEEproof}
Note that the integrand $\varepsilon (f) + \varepsilon (-f)$ in (\ref{problem: inner}) can be rewritten as
\vspace{-6mm}
\begin{equation}\label{eq: MSE_pos_neg}
\varepsilon (f) + \varepsilon (-f) =  T \frac{M(-fT)(1+c(f)\bar{k}(fT)) + M(fT)(1 + c(-f)\bar{k}(fT)) }{ 1+c(f) + c(-f) + c(f) c(-f)\bar{k}(fT)}, \vspace{-5mm}
\end{equation}
where $\bar{k}(f) \triangleq 1-k(f)^2$.
Since $\varepsilon (f) + \varepsilon (-f)$ evaluated at some $f_0$ is a function only of $\bm{s}(f_0)$ and $\bm{s}(-f_0)$ through $c(f_0)$ and $c(-f_0)$, respectively, we just need to minimize by optimizing $c(f_0)$ and $c(-f_0)$ in the integrand at each $f\in{\mathcal{F}^+}$ subject to the constraint.
Let $c(f_0)=c_1$ and $c(-f_0)=c_2$.
Then, it can be shown that $\partial \big(\varepsilon (f_0) + \varepsilon (-f_0) \big)/\partial c_1 < 0$ and $\partial \big(\varepsilon (f_0) + \varepsilon (-f_0) \big)/\partial c_2 < 0$.
Moreover, since $a(f_0)=\bm{s}(f_0)^{\mathcal{H}}\bm{s}(f_0)$, $c(f_0)$ is constrained by $a(f_0)$ through $\bm{s}(f_0)$ and $c(-f_0)$ is constrained by $a(-f_0)$ through $\bm{s}(-f_0)$.
Thus, we now can separately find $\bm{s}(f_0)$ that maximizes $c(f_0)$ for given $a(f_0)$ and $\bm{s}(-f_0)$ that maximizes $c(-f_0)$ for given $a(-f_0)$.
This maximization of $c(f)$ defined in (\ref{eq: c_f}) subject to $a(f)=\| \bm{s}(f) \|^2$ is exactly the same problem solved in \cite[Section~IV-B]{Cho_04}, where the optimal solution is given by (\ref{eq: opt_TX_waveform}) at each $f\in {\mathcal{F}}$.
Therefore, the conclusion follows.
\end{IEEEproof}

According to (\ref{eq: opt_TX_waveform}), the optimal $\bm{s}(f)$ given $a(f)$ is not affected by the impropriety frequency function $k(f)$.
However, it actually affects the outer optimization of $a(f)$, which will be performed in what follows.
Let $\lambda(f)$ denote the largest eigenvalue of $\bm{H}(f)^{\mathcal{H}}\bm{R}_N(f)^{-1}\bm{H}(f)$.
Then, by (\ref{eq: opt_TX_waveform}), $c(f)$ can be simplified as $c(f)  = M(fT) \lambda(f) a(f)/T$, $\forall f\in {\mathcal{F}}$.
Thus, the outer minimization problem of Problem~\ref{problem: FD_TX} to find the optimal energy density $a_{\rm opt}(f)$ for $f\in {\mathcal{F}}$ becomes
\begin{problem}\label{problem: Tx_power}
\end{problem}
\vspace{-27mm}
\begin{IEEEeqnarray}{ll}
\underset{a(f),\; a(-f)}{\text{minimize}}&    T^2 \int_{{\mathcal{F}^+}} \bar{\varepsilon}  (f) df\IEEEeqnarraynumspace\IEEEyesnumber\IEEEyessubnumber\label{eq: problem_tx_power1}\\
\,\text{subject to}\;\;&  \frac{1}{T}\int_{{\mathcal{F}^+}} M(fT) a(f) + M(-fT) a(-f)df = P_T, \IEEEeqnarraynumspace\IEEEyessubnumber \label{eq: power constraint_tx_power2}\vspace{-7mm}
\end{IEEEeqnarray}
where $\bar{\varepsilon}(f)$ is given by
\vspace{-5mm}
\begin{equation}
\bar{\varepsilon}(f)  \triangleq \frac{\frac{M(-fT)}{T} \left(1+\frac{M(fT)}{T}\lambda(f) a(f)\bar{k}(fT)   \right) +\frac{M(fT)}{T} \left(1+\frac{M(-fT)}{T}\lambda(-f) a(-f)\bar{k}(fT)   \right)   }{1 + \frac{M(fT)}{T}\lambda(f) a(f) + \frac{M(-fT)}{T}\lambda(-f) a(-f) + \frac{M(fT)}{T}\lambda(f) a(f)  \frac{M(-fT)}{T}\lambda(-f) a(-f) \bar{k}(fT)  },\vspace{-4mm}
\end{equation}
with $\bar{k}(f) \triangleq 1-k(f)^2$ as already used in (\ref{eq: MSE_pos_neg}).
Now, we are ready to present the optimal $a(f)$.
In what follows, ${\mathcal{F}}_M$ and ${\mathcal{F}}_{\lambda}$ denote the supports of $M(fT)$ and $\lambda(f)$, respectively, i.e., ${\mathcal{F}}_M \triangleq \{f \in {\mathcal{F}}: M(fT) \neq 0  \}$ and ${\mathcal{F}}_{\lambda} \triangleq \{f \in {\mathcal{F}}: \lambda(f) \neq 0  \}$.

\begin{proposition}\label{proposition: optimize_power}
The optimal solution to Problem~\ref{problem: Tx_power} can be found by performing a line search for a parameter $\nu$ in $(0, \; \nu_{\max}]$, where $\nu_{\max}\triangleq \max_f \lambda(f)\big( M(-fT) k(fT)^2 + M(fT) \big)$.
For each $\nu\in(0, \; \nu_{\max}]$, a candidate density function can be constructed by using the algorithm described in Table~\ref{table: 1}, where
\vspace{-4mm}
\begin{equation} \label{eq: optimal_a_f_update}
u(\nu, f) = \left\{ \begin{array}{ll}
\,\vspace{-0.70in}\\
\displaystyle \left[ \frac{1}{\sqrt{\nu}} - \frac{1+ M(-fT) \lambda(-f) a(-f) }{\sqrt{\lambda(f)\nu(f)}} \right]^+
\frac{\sqrt{\lambda(f)\nu(f) }}{ \lambda(f) M(fT)  g(-f) },  &  \text{ for }  f\in  {\mathcal{F}}_M \cap {\mathcal{F}}_{\lambda}, \\
\,\vspace{-0.75in}\\
0, &  \text{ for }  f\in  {\mathcal{F}}_M \cap ({\mathcal{F}}_{\lambda})^c, \\
\,\vspace{-0.75in} \\
\text{arbitrary}, & \text{ for } f\in  ({\mathcal{F}}_M)^c,  \,\vspace{-0.15in}
\end{array} \right. \vspace{-3mm}
\end{equation}
with $g(f) \triangleq 1 + M(fT) \lambda(f) a(f) \bar{k}(fT)$, $\nu(f) \triangleq M(-fT) k(fT)^2 + M(fT)g(-f)^2$, and $[x]^+ \triangleq \max (x, 0)$.
The candidate function that satisfies the power constraint (\ref{eq: power constraint_tx_power2}) is the optimal density function $a_\text{opt}(f)$.
\end{proposition}

\begin{IEEEproof}
See Appendix~C.
\end{IEEEproof}


Note that any line search algorithm can be used to find $a_\text{opt}(f)$ in Proposition~\ref{proposition: optimize_power}.
Note also that the algorithm in Table~\ref{table: 1} allows the construction of an approximate solution with arbitrary accuracy if the interval $\mathcal{F}^+$ is partitioned finely enough.

Now, by using $a_\text{opt}(f)$, we can find the VFTs of the optimal transmit and receive waveforms as follows.

\begin{theorem}\label{thm: 1}
The VFT $\bm{s}_{\rm opt}(f)$ of the jointly optimal transmit waveform $s_{\rm opt}(t)$ as the solution to Problem~\ref{problem: 1} are given by
\vspace{-7mm}
\begin{IEEEeqnarray}{rCl}
\bm{s}_{\rm opt}(f) &=& \left\{\begin{array}{ll}
\,\vspace{-0.80in}\\
\sqrt{a_{\rm opt}(f)} \bm{v}(f)e^{{\rm j} \theta(f)}, & \text{for } f\in  {\mathcal{F}}_M,
\,\vspace{-0.30in}\\
\text{arbitrary}, & \text{for } f\in ( {\mathcal{F}}_M )^c, \,\vspace{-0.15in}\\
\end{array} \right. \IEEEeqnarraynumspace\IEEEyesnumber\label{eq: opt_tx_final}\vspace{-5mm}
\end{IEEEeqnarray}
where $\theta(f)$ can be chosen arbitrarily.
Then, the VFTs $\bm{w}_{1,{\rm opt}}(f)$ and $\bm{w}_{2,{\rm opt}}(f)$ of the jointly optimal receive waveforms $w_{1, {\rm opt}}(t)$ and $w_{2, {\rm opt}}(t)$ can be found by using (\ref{eq: Rx_opt}).
\end{theorem}

\begin{IEEEproof}
The conclusion immediately follows from the relation (\ref{eq: def_augmented_VFT}) among $\bm{s}_{\rm opt}(f)$, $\bar{\bm{s}}_{\rm opt}(f)$, $\bm{w}_{1,{\rm opt}}(f)$, $\bm{w}_{2,{\rm opt}}(f)$, and $\bar{\bm{w}}_{\rm opt}(f)$, and Propositions~\ref{proposition: inner_optimization} and \ref{proposition: optimize_power}.
\end{IEEEproof}

As already mentioned, cyclostationarity and impropriety, respectively, imply the periodic spectral correlation and the symmetric spectral correlation about the origin \cite[Ch.~10]{Schreier_10}, \cite{Yeo_14}.
Theorem~1 vividly shows these structures in the optimal transmitted signal.
Specifically, the use of the VFT technique and the augmentation of $s(f)$ and $s(-f)$ to form $\bar{s}(f)$ take care of the periodic spectral correlation and the symmetric spectral correlation, respectively.

\section{Numerical Results}

In this section, numerical results are provided that show the magnitude square of the optimal transmit and receive waveforms and that show the MSE performance achieved by the optimal waveforms as a function of the amount of impropriety.
For illustrative purposes, it is assumed throughout this section that an interferer linearly modulates a data sequence consisting of  uncorrelated zero-mean proper-complex QPSK symbols with $E_s/ N_0 = 10$ [dB] and a square-root raised cosine transmit waveform having excess bandwidth $\beta=0.25$.
It is assumed that the Tx linearly modulates a data sequence consisting of uncorrelated zero-mean improper-complex QAM symbols with uncorrelated in-phase and quadrature components.
It is also assumed that all the channels are frequency flat and corrupted by AWGN.

\begin{figure}[tbp]\centering
\includegraphics[width=6in]{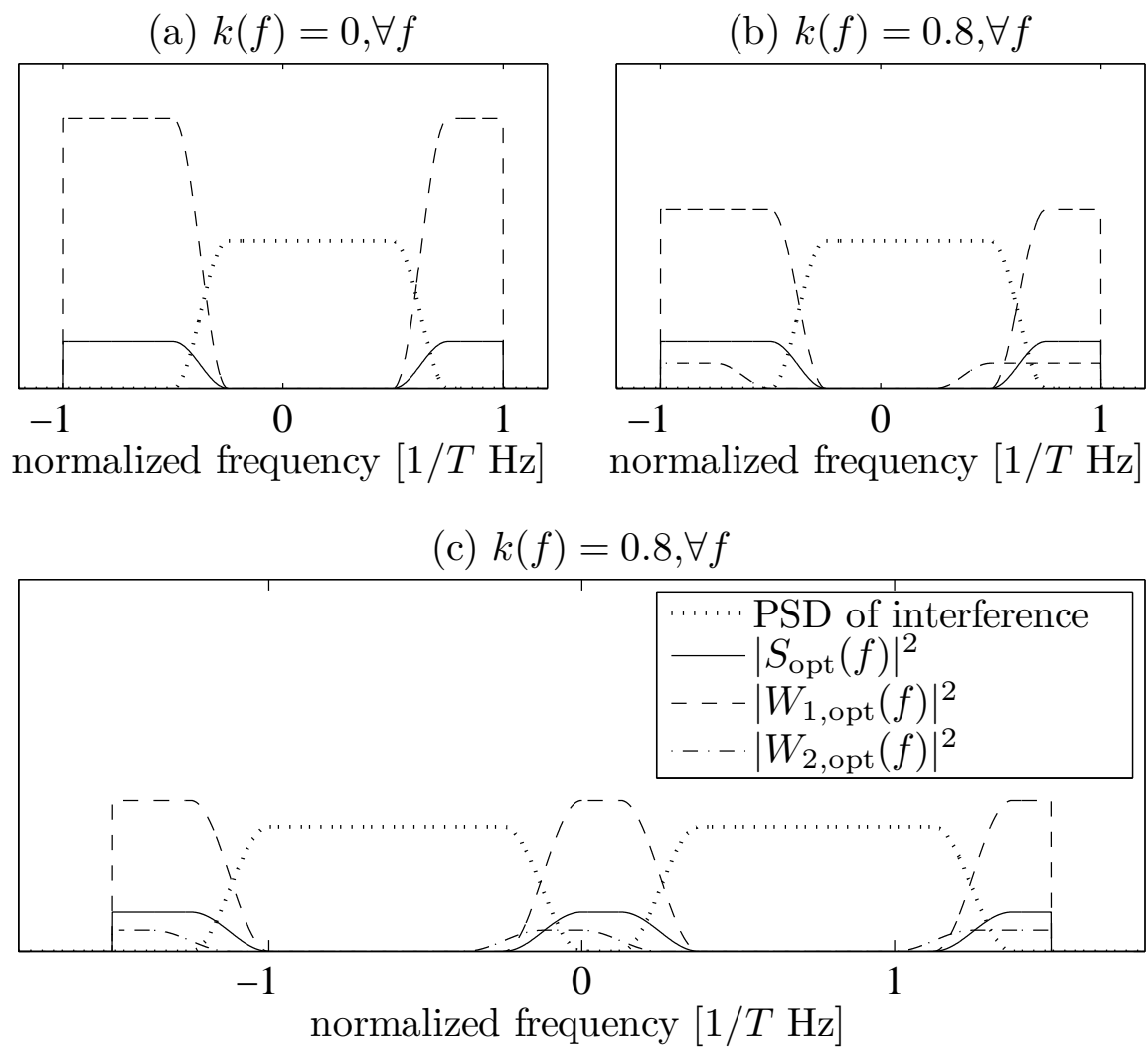} %
\caption{Comparison of squared-magnitudes of the optimal transmit and receive waveforms for (a) $k(f)=0,\forall f$ and single interferer, (b) $k(f)=0.8,\forall f$ and single interferer, and (c) $k(f)=0.8,\forall f$ and two uncorrelated interferers.} \label{Fig: spectrum}
\end{figure}

The first results are to compare the PSD of data-like interference with the squared magnitudes of the optimal transmit and receive waveforms.
There is a single interferer in Figs.~\ref{Fig: spectrum}-(a) and (b), whereas there are two uncorrelated interferers in Fig.~\ref{Fig: spectrum}-(c).
The QAM symbols of the Tx have $E_s/N_0 = 5$ [dB].
For Fig.~\ref{Fig: spectrum}-(a), the QAM symbols have the in-phase variance the same as the quadrature variance, which implies $k(f)=0, \forall f$.
For Figs.~\ref{Fig: spectrum}-(b) and (c), the QAM symbols have the in-phase variance $4$-times the quadrature variance, which implies $k(f)=0.8, \forall f$.
It can be seen that $w_{2, {\rm opt}}(t)$, processing the complex conjugate of the received signal, is zero for the data sequence having $k(f)=0, \forall f$, but it is non-zero for the data sequence having $k(f)=0.8, \forall f$.

\begin{figure}[tbp]\centering
\includegraphics[width=6in]{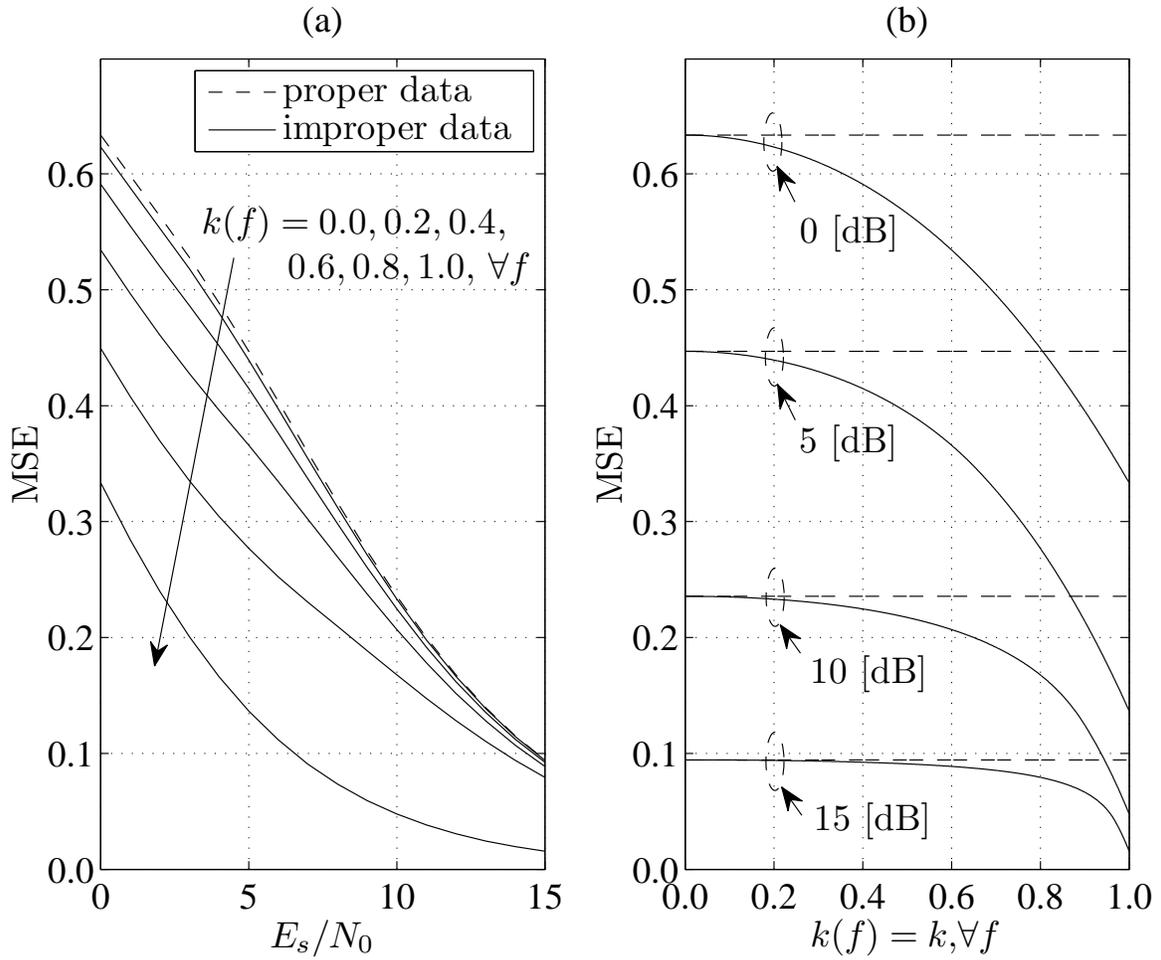} %
\caption{Comparison of MSE (a) versus $E_s/N_0$ and (b) versus impropriety $k(f)=k,\forall f$.} \label{Fig: MSE}
\end{figure}

The next results are to compare the MSEs achieved by the optimal transmit and receive waveforms for different levels of impropriety.
We consider the same number of interferers and interference parameters as Fig.~\ref{Fig: spectrum}-(c).
In Fig.~\ref{Fig: MSE}-(a), the QAM symbols of the Tx have $E_s/N_0 $ from $0$ to $15$ [dB] and have $k(f)=0.0, 0.2, 0.4, 0.6, 0.8$, or $1.0, \forall f$.
In Fig.~\ref{Fig: MSE}-(b), the QAM symbols of the Tx have  $E_s/N_0 =0, 5, 10$ or $15$ [dB] and have $k(f)$ from $0$ to $1, \forall f$.
In both cases, as the amount of impropriety increases, the optimal pair of the Tx and Rx more exploits impropriety and cyclostationarity of the desired signal in suppressing the data-like interference and, consequently, the MSE performance monotonically improves.


\section{Conclusions}
In this paper, we have considered a joint optimization of the Tx and Rx for the transmission of an improper-complex SOS data sequence over an additive proper-complex cyclostationary noise channel.
An MSE minimization problem is formulated under the average transmit power constraint to find the jointly optimal transmit waveform of a linear modulator and the receive waveforms of a widely linear Rx.
This problem is converted into an equivalent problem described in the frequency domain with the help of the VFT technique and solved by introducing the notion of the impropriety frequency function.
It is shown that the optimal transmit and receive waveforms well exploit the frequency-domain second-order structure of the improper-complex SOS data sequence and the additive proper-complex SOCS noise.

\section*{Appendix}

%

\subsection{Proof of {Lemma~\ref{lemma: MSE_Rx}}}

\begin{IEEEproof}
Define the $2$-by-$2$ matrices $\hat{\bm{M}}(f)$, $\bm{M}(f)$, and $\bm{K}(f)$, respectively, as
\vspace{-6mm}
\begin{IEEEeqnarray}{l}
 \hat{\bm{M}}(f) \triangleq \begin{bmatrix}
\,\vspace{-0.75in}\\
M(f) &\!\! \tilde{M}(f) \\
\,\vspace{-0.80in}\\
\tilde{M}(f)^*& \!\! M(-f)^* \,\vspace{-0.1in}\\
\end{bmatrix}\!\!, \; \bm{M}(f) \triangleq \begin{bmatrix}
\,\vspace{-0.75in}\\
M(f)e^{{\rm j} \phi(f)} & 0 \\
\,\vspace{-0.80in}\\
0 &\!\!\!\! M(-f)e^{-{\rm j} \phi(f)} \,\vspace{-0.1in}\\
\end{bmatrix}^{\frac{1}{2}}\!\!, \; \text{and} \; \bm{K}(f) \triangleq \begin{bmatrix}
\,\vspace{-0.75in}\\
1 &\!\! k(f) \\
\,\vspace{-0.80in}\\
k(f) &\!\! 1 \,\vspace{-0.1in}\\
\end{bmatrix}\!\!.\vspace{-6mm} \IEEEeqnarraynumspace\IEEEyesnumber
\end{IEEEeqnarray}
Then, we can rewrite $\hat{\bm{M}}(f)$ as $\hat{\bm{M}}(f) = \bm{M}(f)\bm{K}(f)\bm{M}(f)^{\mathcal{H}}$.
Also, define the $\bar{\mathcal{N}}(f)$-by-$\bar{\mathcal{N}}(f)$ matrix $\bar{\bm{R}}_N(f)$ and the $\bar{\mathcal{N}}(f)$-by-$2$ matrix $\bar{\bm{S}}(f)$ as $\bar{\bm{R}}_N(f)\triangleq \text{diag}\big\{\bm{R}_N(f), \bm{J}(-f)\bm{R}_N(-f)^*\bm{J}(-f)\big\}$ and $\bar{\bm{S}}(f) \triangleq \text{diag}\big\{\bm{s}(f), \bm{J}(-f) \bm{s}(-f)^*\big\}$, respectively.
Due to the ambient noise component in $N(t)$, $\bm{R}_N(f)$ and $\bar{\bm{R}}_N(f)$ are positive definite for all $f\in{\mathcal{F}}$.
By using $\bar{\bm{R}}_N(f)^{-1/2}$, define the $\bar{\mathcal{N}}(f)$-by-$2$ matrix $\tilde{\bm{P}}(f)$ as $\tilde{\bm{P}}(f) \triangleq \bar{\bm{R}}_N(f)^{-1/2}\bar{\bm{H}}(f)\bar{\bm{S}}(f)\bm{M}(fT)/ \sqrt{T}$.
Then, it can be shown that $\bar{\bm{H}}(f) \bar{\bm{M}}(fT)$ $\bar{\bm{s}}(f)=  \sqrt{T M(fT)}e^{-{\rm j} \phi(fT)/2}$ $ \bar{\bm{R}}_N(f)^{1/2} \tilde{\bm{P}}(f) \bm{k}(fT)$.
Thus, the second term of the integrand in (\ref{eq: Rx_opt_MSE}), which contains $\bar{\bm{H}}(f) \bar{\bm{M}}(fT)\bar{\bm{s}}(f)$, can be rewritten as
\vspace{-8mm}
\begin{IEEEeqnarray}{rCl}
&&\bar{\bm{s}}(f) ^{\mathcal{H}}  \bar{\bm{M}}(fT)^{\mathcal{H}} \bar{\bm{H}}(f)^{\mathcal{H}}  \bar{\bm{R}}(f)^{-1} \bar{\bm{H}}(f) \bar{\bm{M}}(fT) \bar{\bm{s}}(f)  \nonumber\\
&=& T M(fT)\bm{k}(fT)^{\mathcal{T}}\tilde{\bm{P}}(f)^{\mathcal{H}}  \Big(  \bm{I} +\tilde{\bm{P}}(f)\bm{K}(fT)  \tilde{\bm{P}}(f)^{\mathcal{H}}  \Big)^{-1} \tilde{\bm{P}}(f)\bm{k}(fT),  \;\;\;\IEEEeqnarraynumspace\IEEEyesnumber\label{eq: prp_1}\vspace{-8mm}
\end{IEEEeqnarray}
where $\bm{I}$ conveniently denotes the appropriately sized identity matrix throughout this proof.
Let $\hat{\bm{p}}(f) \triangleq \sqrt{M(fT) /T} e^{{\rm j} \phi(f)/2} \bm{R}_N(f)^{-1/2} \bm{H}(f) \bm{s}(f)  $.
Then, we can rewrite $\tilde{\bm{P}}(f)$ and $c(f)$ defined in (\ref{eq: c_f}) as $\tilde{\bm{P}}(f) = \text{diag}\big\{ \hat{\bm{p}}(f),$ $\bm{J}(-f) \hat{\bm{p}}(-f)^*\big\}$ and $c(f)= \|\hat{\bm{p}}(f)\|^2$, respectively.
If $c(f)c(-f) = 0$, then it can be shown that (\ref{eq: prp_1}) leads to (\ref{eq: MSE_simple}) by using the matrix inversion lemma showing $ \bm{I} - \bm{u}^{\mathcal{H}} (\bm{I} + \bm{u}\bm{u}^{\mathcal{H}})^{-1} \bm{u}=( 1 + \bm{u}^{\mathcal{H}}\bm{u})^{-1}$ for any vector $\bm{u}$.
If $c(f)c(-f) \neq 0$, then, since $\tilde{\bm{P}}(f)^{\mathcal{H}} \tilde{\bm{P}}(f) = \text{diag}\big\{c(f), c(-f)\big\}$ is invertible, it can be shown that $\tilde{\bm{P}}(f)^{\mathcal{H}}  \big( \bm{I} +\tilde{\bm{P}}(f)\bm{K}(fT)  \tilde{\bm{P}}(f)^{\mathcal{H}}  \big)^{-1} \tilde{\bm{P}}(f)\tilde{\bm{C}}(f) = \bm{I}$, where $\tilde{\bm{C}}(f)$ is defined as $\tilde{\bm{C}}(f)\triangleq \big( \tilde{\bm{P}}(f)^{\mathcal{H}} \tilde{\bm{P}}(f) \big)^{-1} +  \bm{K}(fT)$.
Since $c(f)$ and $c(-f)$ are not zero, we can rewrite $\tilde{\bm{C}}(f)$ as $\tilde{\bm{C}}(f) = \text{diag}\big\{c(f)^{-1} ,$ $c(-f)^{-1} + 1 -  k(fT)^2 \big\} + \bm{k}(fT)\bm{k}(fT)^{\mathcal{T}}$.
Thus, we now can rewrite the right side of (\ref{eq: prp_1}) as $ T M(fT)\bm{k}(fT)^{\mathcal{T}} \tilde{\bm{C}}(f)^{-1}  \bm{k}(fT)$.
By using the matrix inversion lemma, the conclusion follows.
\end{IEEEproof}
\subsection{Proof of {Proposition~\ref{proposition: optimize_power}}}
\begin{IEEEproof}
For convenience, the integration interval ${\mathcal{F}^+}$ is partitioned into $N$ equal-length subintervals.
Then, the solution can be straightforwardly extended to the original problem by letting $N$ tend to infinity.
Let $\xi_i \triangleq i/(2NT) - 1/(4NT)$, $a_i \triangleq a(\xi_i)$, $\hat{a}_i \triangleq a(-\xi_i)$, $m_i \triangleq M(\xi_i T)/T$, $\hat{m}_i \triangleq M(-\xi_i T)/T$, $\lambda_i \triangleq \lambda(\xi_i)$, $\hat{\lambda}_i \triangleq \lambda(-\xi_i)$, and $k_i \triangleq k(\xi_i T)$.
Then, the original optimization problem can be approximated by
\vspace{-5mm}
\begin{IEEEeqnarray}{C}\label{eq: DT opt problem}
\underset{a_i, \hat{a}_i \geq 0}{\text{minimize}} \quad   \sum_{i=1}^{N} f_i(a_i, \hat{a}_i) \quad \text{subject to} \quad \sum_{i=1}^{N} \big(m_i a_i + \hat{m}_i \hat{a}_i\big)  \leq P_T T,  \IEEEeqnarraynumspace\IEEEyesnumber \vspace{-5mm}
\end{IEEEeqnarray}
where $f_i(a_i, \hat{a}_i)$ is given by $f_i(a_i, \hat{a}_i) \triangleq \big( \hat{m}_i (1+ m_i \lambda_i a_i \bar{k}_i   ) + m_i (1+\hat{m}_i \hat{\lambda}_i \hat{a}_i \bar{k}_i   )   \big)/ \big(1 + m_i \lambda_i a_i + \hat{m}_i \hat{\lambda}_i \hat{a}_i + m_i \lambda_i a_i  \hat{m}_i\hat{\lambda}_i \hat{a}_i \bar{k}_i \big)$
for non-negative real numbers $m_i, \hat{m}_i,  \lambda_i, \hat{\lambda}_i$, and $\bar{k}_i \triangleq 1 - k_i^2$ with $ 0 \leq k_i <1$, $\forall i$.
It can be easily shown that, if $m_i=0$, $a_i$ can be chosen arbitrarily because $a_i$ does not affect both the objective function and the constraint.
It can be also easily shown that $\lambda_i=0$ results in $a_i=0$ to keep from wasting the transmit power.
Similarly, if $\hat{m}_i=0$, then $\hat{a}_i$ can be chosen arbitrarily, and if $\hat{\lambda}_i=0$, then $\hat{a}_i=0$.
The case of $k_i = 1$ is discussed after solving the optimization problem for $k_i < 1$.
Thus, in what follows, we assume that $m_i \hat{m}_i\neq 0$, $\lambda_i \hat{\lambda}_i\neq 0$, and $k_i < 1$, $\forall i$.

Define $\bm{a}$ and $\bm{m}$ as $\bm{a} \triangleq [a_1, \hat{a}_1, a_2, \hat{a}_2, \cdots, a_N, \hat{a}_N]^{\mathcal{T}}$ and $\bm{m} \triangleq [m_1, \hat{m}_1, m_2, \hat{m}_2, \cdots, m_N, \hat{m}_N]^{\mathcal{T}}$, respectively.
Then, it can be easily shown that the Hessian $\bm{F}_i(\bm{a})$ of the objective function $\sum_{i=1}^{N} f_i(a_i, \hat{a}_i)$ is a positive definite matrix for each $\bm{a}$ and the equality constraint $\sum_{i=1}^{N} \big(m_i a_i + \hat{m}_i \hat{a}_i\big)  = \bm{m}^{\mathcal{T}}\bm{a}$ is an affine function of $\bm{a}$.
Thus, the problem in (\ref{eq: DT opt problem}) is a strictly convex optimization problem.
Since the Karush-Kuhn-Tucker (KKT) condition is necessary and sufficient for a point to be the unique solution of a strictly convex optimization problem \cite[Theorem~22.9]{Chong_13}, we first need to find the KKT condition.

The Lagrangian function of (\ref{eq: DT opt problem}) can be written as $l(\bm{a}, \nu, \bm{u}) =  \sum_{i=1}^{N} f_i(a_i, \hat{a}_i) + \nu ( \bm{m}^{\mathcal{T}}\bm{a} - P_T T) - \bm{\mu}^{\mathcal{T}}\bm{a}$ by introducing the multipliers $\nu$ and $\bm{\mu} \triangleq [\mu_1, \hat{\mu}_1, \mu_2, \hat{\mu}_2, \cdots, \mu_N, \hat{\mu}_N]^{\mathcal{T}}$.
Then, the KKT condition can be written as $-m_i \lambda_i \big( \hat{m}_i k^2 + m_i \hat{g}_i(\hat{a}_i)^2 \big)/{h_i(a_i,\hat{a}_i)^2}  + m_i \nu - \mu_i  = 0$ and $-\hat{m}_i \hat{\lambda}_i  \big( m_i k^2 + \hat{m}_i g_i(a_i) ^2 \big) /{h_i(a_i,\hat{a}_i)^2} + \hat{m}_i \nu - \hat{\mu}_i = 0$ with $a_i \geq  0$, $\hat{a}_i \geq  0$, $\mu_i \geq  0$, $\hat{\mu}_i \geq 0$, $\mu_i a_i = 0$, $\hat{\mu}_i \hat{a}_i=0$, $\forall i$, and $\sum_{i=1}^{N} (m_i a_i + \hat{m}_i \hat{a}_i) = P_T T$, where $h_i(a_i,\hat{a}_i) \triangleq 1 + m_i \lambda_i a_i + \hat{m}_i \hat{\lambda}_i \hat{a}_i + m_i \lambda_i a_i  \hat{m}_i\hat{\lambda}_i \hat{a}_i \bar{k}_i $, $g_i(a_i) \triangleq 1+ m_i \lambda_i a_i \bar{k}_i$, and $\hat{g}_i(\hat{a}_i)  \triangleq 1+\hat{m}_i \hat{\lambda}_i \hat{a}_i \bar{k}_i  $.

Define $\nu_i (a_i, \hat{a}_i) \triangleq \lambda_i \big( \hat{m}_i k^2 + m_i \hat{g}_i(\hat{a}_i)^2 \big)/ h_i(a_i,\hat{a}_i)^2$ and $\hat{\nu}_i (a_i, \hat{a}_i) \triangleq  \hat{\lambda}_i \big( m_i k^2 + \hat{m}_i g_i(a_i) ^2 \big)/h_i(a_i,\hat{a}_i)^2$, respectively.
It can be easily shown that $\partial \nu_i (a_i, \hat{a}_i) / \partial a_i < 0$, $\partial \nu_i (0, \hat{a}_i)/ \partial \hat{a}_i < 0$, $\partial \hat{\nu}_i (a_i, \hat{a}_i) / \partial \hat{a}_i < 0$, and $\partial \hat{\nu}_i (a_i, 0) / \partial a_i < 0$ for all $a_i \geq 0$ and $\hat{a}_i \geq 0$.
Thus, $\nu_i (a_i, \hat{a}_i) < \nu_i (0, 0)$ and $\hat{\nu}_i (a_i, \hat{a}_i) < \hat{\nu}_i (0, 0)$, respectively, for all $a_i > 0$ and $\hat{a}_i > 0$.
It can be also shown that, if $\nu \geq \nu_i (0, 0)$ and $\nu \geq \hat{\nu}_i (0, 0)$, then only $a_i = 0$ and $\hat{a}_i = 0$ satisfy the KKT condition.
It is noteworthy that $\nu$ satisfying the KKT condition is upper-bounded by $\nu_{\max}$ that is defined as the largest value among $\nu_i (0, 0)$ and $\hat{\nu}_i (0, 0)$, $\forall i$, which can be easily found and is finite and positive.
Thus, to find $\bm{a}$, $\nu$, and $\bm{\mu}$ that jointly satisfy the KKT condition, a line search for $\nu$ can be performed over the interval $(0, \; \nu_{\max}]$, where two steps are needed to construct a candidate solution $\bm{a}$ and the multiplier $\bm{\mu}$ at each $\nu$.

First, a candidate solution $\bm{a}$ associated with $\nu$ is constructed as follows.
Given $\nu$, we need to find the pair of $(a_i,\hat{a}_i)$ satisfying the KKT condition, i.e., $\nu - \mu_i / m_i= \nu_i (a_i, \hat{a}_i)$ and $\nu - \hat{\mu}_i / \hat{m}_i= \hat{\nu}_i (a_i, \hat{a}_i)$ with $a_i \geq  0$, $\hat{a}_i \geq  0$, $\mu_i \geq  0$, $\hat{\mu}_i \geq 0$, $\mu_i a_i = 0$, $\hat{\mu}_i \hat{a}_i=0$, which can be rewritten as
\vspace{-6mm}
\begin{IEEEeqnarray}{rCCCl}
a_i &=& u_1(\hat{a}_i) &\triangleq&  \left[ \sqrt{\lambda_i \big(\hat{m}_i k^2 + m_i \hat{g}_i(\hat{a}_i)^2 \big)}/ \sqrt{\nu} - (1+ \hat{m}_i \hat{\lambda}_i \hat{a}_i)\right]^+  \big(\lambda_i m_i \hat{g}_i(\hat{a}_i)\big)^{-1}, \IEEEeqnarraynumspace\IEEEyesnumber \IEEEyessubnumber \label{eq: KKT_1_update}\\
\hat{a}_i &=& u_2(a_i) &\triangleq&  \left[ \sqrt{ \hat{\lambda}_i \big( m_i k^2 + \hat{m}_ig (a_i)^2 \big)}/\sqrt{\nu} - (1 + m_i \lambda_i  a_i)\right]^+\big(\hat{\lambda}_i \hat{m}_i g (a_i) \big)^{-1}, \IEEEeqnarraynumspace \IEEEyessubnumber \label{eq: KKT_2_update}\vspace{-7mm}
\end{IEEEeqnarray}
$\mu_i =0 $ if $a_i > 0$, and $\hat{\mu}_i =0 $ if $\hat{a}_i > 0$.
It can be easily shown that $u_1(\hat{a}_i)$ is a decreasing function of $\hat{a}_i$ and $u_2(a_i)$ is a decreasing function of $a_i$.
Thus, $(u_2\circ u_1) ( \hat{a}_i)$ becomes an increasing function of $\hat{a}_i$.
It is noteworthy that the non-negative numbers $a_i$ and $\hat{a}_i$ are upper-bounded by $u_1(0)$ and $u_2(0)$.
Thus, when we alternately update $a_i$ and $\hat{a}_i$ from $\hat{a}_i=0$  by using (\ref{eq: KKT_1_update}) and (\ref{eq: KKT_2_update}), respectively, both $a_i$ and $\hat{a}_i$ converge to the solution satisfying the KKT conditions.
This iteration algorithm can be also used to find the candidate solution $a_i$ and $\hat{a}_i$ for the case of $k_i=1$.
Note that, if $k_i=1$ and $\lambda_i=\hat{\lambda}_i$, any pair of $a_i$ and $\hat{a}_i$ satisfying $m_i a_i + \hat{m}_i \hat{a}_i = [ \sqrt{(m_i + \hat{m}_i)/(\lambda_i \nu)} - 1/ \lambda_i ]^+$ can be the candidate solution associated with $\nu$.
After finding $a_i$ and $\hat{a}_i$, $\mu_i$ and $\hat{\mu}_i$ can be computed by substituting $a_i$, $\hat{a}_i$, and $\nu$ into the KKT condition.

Second, after constructing the candidate solution $\bm{a}$ associated with $\nu$, we check whether the candidate solution satisfies the power constraint $\sum_{i=1}^{N} \big(m_i a_i + \hat{m}_i \hat{a}_i\big) = P_T T$.
If so, then the candidate solution associated with $\nu_{\rm opt}$ is the optimal solution $\bm{a}_{\rm opt}$.
If not, then the line search continues.
Therefore, the conclusion follows.
\end{IEEEproof}

\newpage

\begin{table}
\caption{An Algorithm to Construct Candidate Density Function at $\nu\in (0, \; \nu_{\max}]$}\label{table: 1}\vspace*{-0.1in}
\hrulefill\\
\,\vspace{-2mm}
\begin{itemize}
\item[1: ] Choose $f_0\in\mathcal{F}^+$.
\item[2: ] Construct $a(f_0)$ and $a(-f_0)$ as follows.
\item[3: ] \begin{itemize} \item[] $\quad$ Set $a(-f_0):=0$. \end{itemize}
\item[4: ] \begin{itemize} \item[] $\quad$ \textbf{REPEAT} \end{itemize}
\item[5: ] \begin{itemize} \item[] $\quad$ Update $a(f_0)$ as $a(f_0) := u(\nu, f_0)$ by using $u(\nu, f)$ defined in (\ref{eq: optimal_a_f_update}).\end{itemize}
\item[6: ] \begin{itemize} \item[] $\quad$ Update $a(-f_0)$ as $a(-f_0):= u(\nu, -f_0)$.\end{itemize}
\item[7: ] \begin{itemize} \item[] $\quad$ \textbf{UNTIL} $a(f_0)$ and $a(-f_0)$ converge.\end{itemize}
\item[8: ] Repeat lines $1-7$ for all $f_0\in\mathcal{F}^+$.
\end{itemize}
\hrulefill
\end{table}

%

\end{document}